\def   \ni {\noindent}

\def   \bsk {\vskip 15truept}

\def   \newline {\hfil\break}

\documentstyle[epsfig]{article}
\begin{document}

%
% A&A definitions
\def\la{\mathrel{\mathchoice {\vcenter{\offinterlineskip\halign{\hfil
$\displaystyle##$\hfil\cr<\cr\sim\cr}}}
{\vcenter{\offinterlineskip\halign{\hfil$\textstyle##$\hfil\cr
<\cr\sim\cr}}}
{\vcenter{\offinterlineskip\halign{\hfil$\scriptstyle##$\hfil\cr
<\cr\sim\cr}}}
{\vcenter{\offinterlineskip\halign{\hfil$\scriptscriptstyle##$\hfil\cr
<\cr\sim\cr}}}}}
\def\ga{\mathrel{\mathchoice {\vcenter{\offinterlineskip\halign{\hfil
$\displaystyle##$\hfil\cr>\cr\sim\cr}}}
{\vcenter{\offinterlineskip\halign{\hfil$\textstyle##$\hfil\cr
>\cr\sim\cr}}}
{\vcenter{\offinterlineskip\halign{\hfil$\scriptstyle##$\hfil\cr
>\cr\sim\cr}}}
{\vcenter{\offinterlineskip\halign{\hfil$\scriptscriptstyle##$\hfil\cr
>\cr\sim\cr}}}}}
\def\degr{\hbox{$^\circ$}}
\def\arcmin{\hbox{$^\prime$}}
\def\arcsec{\hbox{$^{\prime\prime}$}}

\hsize 5truein
\vsize 8truein
\font\abstract=cmr8
\font\keywords=cmr8
\font\caption=cmr8
\font\references=cmr8
\font\text=cmr10
\font\affiliation=cmssi10
\font\author=cmss10
\font\mc=cmss8
\font\title=cmssbx10 scaled\magstep2
\font\alcit=cmti7 scaled\magstephalf
\font\alcin=cmr6 
\font\ita=cmti8
\font\mma=cmr8
\def\ref{\par\noindent\hangindent 15pt}
\null
%\vskip 3.0truecm
%\baselineskip = 12pt

% beginning of font "title"

\title{\ni Evidence for non-Gaussianity in the CMB}

% beginning of font "author and affiliation"
\bsk \bsk
\author{\ni Jo\~{a}o Magueijo$^1$, Pedro G. Ferreira $^2$, and  
Krzysztof M. G\'orski$^{3,4}$}                                                       
\bsk
\affiliation{\\$^1$Theoretical Physics, 
Imperial College, Prince Consort Road, London SW7 2BZ, UK\\
$^2$Center for Particle Astrophysics, University of California, 
Berkeley, CA94720, USA\\$^3$Theoretical 
Astrophysics Center, Juliane Maries Vej 30,
DK-2100, Copenhagen \O, Denmark\\
$^4$ Warsaw University Observatory, Warsaw, Poland
}                                                
\bsk
\baselineskip = 12pt

% beginning of font "abstract and keywords"
\abstract{ABSTRACT \ni
In a recent Letter we have shown how COBE-DMR maps may be
used to disprove Gaussianity at a high confidence level.
In this report we digress on a few issues closely related
to this Letter. We present the general formalism for surveying
non-Gaussianity employed. We present a few more tests for 
systematics. We wonder about the theoretical implications
of our result. }
\bsk
\baselineskip = 12pt

\bsk
%\baselineskip = 12pt

% beginning of font "text"

\text{

\section{Introduction}
It is hard to overemphasise the importance of Gaussianity in
theories of structure formation. Under the assumption of
Gaussianity calculations become much simpler, and 
only the power spectrum of fluctuations has to be computed.
Furthermore it is thought that Gaussianity is a hallmark of the
inflationary paradigm \cite{inflation}. A simple argument 
for this hinges 
on the fact that small perturbations during inflation satisfy
an harmonic oscillator equation, at the relevant scales
for structure formation. The ground state of an harmonic 
oscillator has a Gaussian wave function. Hence although an
unperturbed background is the classical ground state, quantum
theory forces the existence of ``zero-level'' fluctuations
with Gaussian statistics. These quantum fluctuation are the
seeds of structure, according to inflation, and one identifies
quantum probability with the classical statistics of these
macroscopic seeds. This simple argument linking inflation
and Gaussian statistics is bypassed in non minimal models
of inflation \cite{salopek}.

In a recent Letter \cite{us} we showed how Gaussianity 
could be disproved at a high confidence level, using COBE-DMR maps. 
Here we review that Letter and report on work in progress expanding 
\cite{us}. In Section 2 we describe a generalisation of the  method 
employed in \cite{us}, which incorporates inter-$\ell$
correlators. In Section 3 we review how the method was applied 
to the data in \cite{us}. In Section 4 we review the checks for 
systematics considered in \cite{us} and present a few extra
tests.

We conclude with a few remarks on recent work
\cite{nov,pando,peeb,nvs} which comments on our result.

\section{The statistics} \label{method}
The statistics used in \cite{us} are a subset of a general class
of statistics, to be presented in \cite{usnew}. They are inspired by 
\cite{jm,fermag,fermasi}. The idea is to extract from 
a laboriously produced CMB map  all the relevant information, with
as little theoretical prejudice as possible.
If we believe inflation is the answer, then one may simply reduce a map
to a fit for a set of cosmological parameters. A less omniscient approach 
is to assume only Gaussianity, and concentrate on an unbiased estimate
of the CMB power spectrum (see \cite{gorski97} for an example). 
The ultimate open mind would not assume Gaussianity, but try to extract 
from the map the whole set of correlation functions characterising the 
most general random process. 

It is fair to keep one prejudice: statistical isotropy.
The idea is therefore to extract from a map with $N$ pixels
$N-3$ rotationally invariant independent quantities. We work
in the spherical harmonic representation $a^\ell_m$. 
To construct a an $n$-linear
invariant one takes the tensor product of $n$ $\Delta T_\ell$s
\begin{eqnarray}
(\Delta T_{\ell_1} \otimes \Delta T_{\ell_2} \otimes \cdots \otimes
\Delta T_{\ell_n})({\bf n})&=&\sum_{m_1}\sum_{m_2}\cdots\sum_{m_n}
a_{\ell_1 m_1}a_{\ell_2 m_2}\cdots a_{\ell_n m_n}\nonumber \\ &\times&
Y_{\ell_1 m_1} ({\bf n})\otimes Y_{\ell_2 m_2} ({\bf n})\otimes
\cdots\otimes Y_{\ell_n m_n} ({\bf n})
\end{eqnarray}
One is interested in rotationally invariant quantities. These can be
trivially obtained if one rewrites the tensor product in terms of
the total angular momentum basis. The coefficient of the singlet will
be the higher order invariant we are looking for.

To illustrate the technique we shall work out a few examples. 
Let us first construct the rotationally invariant bilinear. For this
we take the tensor product $\Delta T_{\ell_1} \otimes \Delta T_{\ell_2}$:
\begin{eqnarray}
(\Delta T_{\ell_1} \otimes \Delta T_{\ell_2})({\bf n})
=\sum_{m_1m_2}a_{\ell_1 m_1}a_{\ell_2 m_2}
Y_{\ell_1 m_1} ({\bf n})\otimes Y_{\ell_2 m_2} ({\bf n})
\end{eqnarray}
One can now use the angular momentum addition formulas to
find the the coefficient of the singlet. Defining
\begin{eqnarray}
Y_{\ell_1 m_1} ({\bf n})\otimes Y_{\ell_2 m_2} ({\bf n})=
\sum_{LM}\rangle
\langle LM |\ell \ell m_1 m_2\rangle Y_{LM}({\bf n})
\end{eqnarray}
we want the $L=0$ term. This implies that $\ell_1=\ell_2=\ell$ and from the
condition $M=0$ we have $m_1+m_2=0$. So the coefficient of the singlet
is 
\begin{eqnarray}
I^2_\ell=\sum_{m}a_{\ell
m}a_{\ell -m}
\frac{(-1)^{\ell-m}}{\sqrt{2\ell+1}}
\end{eqnarray}
Up to normalisation this is the simplest quadratic
estimator of the power spectrum.
%; the $(-1)^m$ comes from
%$a_{\ell -m}=(-1)^ma^*_{\ell m}$.

The next simplest case is 
the cubic invariant. Let us first restrict ourselves
to one ``ring'', i.e. fixed $\ell$:
\begin{eqnarray}
(\Delta T_\ell \otimes \Delta T_\ell &\otimes& \Delta T_\ell)({\bf n})
\nonumber \\&=&\sum_{m_1m_2m_3}a_{\ell m_1}a_{\ell m_2}a_{\ell m_3}
Y_{\ell_1 m_1} ({\bf n})\otimes Y_{\ell_2 m_2} ({\bf n})\otimes
Y_{\ell_3 m_3} ({\bf n})
\nonumber \\&=&
\sum_{m_1m_2m_3}a_{\ell m_1}a_{\ell m_2}a_{\ell m_3}
\sum_{LM} \langle  LM |\ell \ell m_1 m_2\rangle 
Y_{LM}({\bf n})\otimes Y_{\ell
m_3}({\bf n})\nonumber \\ &=&
\sum_{m_1m_2m_3}\sum_{LL'MM'}a_{\ell m_1}a_{\ell m_2}a_{\ell m_3}
 \langle LM|\ell \ell m_1 m_2\rangle
\langle  L'M'|L \ell M m_3\rangle Y_{L'M'}({\bf n})\nonumber 
\end{eqnarray}
Once again, we want the singlet. From $L'=M'=0$ we get $L=\ell$ and
$M=-m_3$. This leaves us with
\begin{eqnarray}
%I^3_\ell&=&
&&\sum_{m_1m_2m_3}a_{\ell m_1}a_{\ell m_2}a_{\ell m_3}
\langle  \ell -m_3|\ell \ell m_1 m_2\rangle
\langle  0 0|\ell \ell -m_3 m_3\rangle \nonumber \\&=&\frac{1}{\sqrt{(2\ell+1)}}
(-1)^{\ell}\sum_{m_1m_2m_3}a_{\ell m_1}a_{\ell m_2}a_{\ell m_3}
\left ( \begin{array}{ccc} \ell & \ell & \ell \\ m_1 & m_2 & m_3
\end{array} \right ) \delta_{m_1+m_2+m_3, 0} \nonumber 
\end{eqnarray}
From the symmetry properties of the Wigner $3J$ coefficients we
immediately see that for $\ell$ odd this quantity is identically zero.
%This is proportional to the estimator used in \cite{letter}. 

We are
also interested in relating power between $\ell$s. Consider then 
$(\Delta T_{\ell-1} \otimes \Delta T_\ell \otimes \Delta T_{\ell+1})({\bf
n})$. From the same manipulations one finds
\begin{eqnarray}
J^3_\ell&=&
\sum_{m_1m_2m_3}a_{\ell-1 m_1}a_{\ell+1 m_2}a_{\ell m_3}
\langle  \ell -m_3|\ell+1 \ell-1 m_1 m_2\rangle
\langle  0 0|\ell \ell -m_3 m_3\rangle \nonumber 
\\&=& \frac{1}{\sqrt{(2\ell+1)}}
(-1)^{\ell}\sum_{m_1m_2m_3}a_{\ell m_1}a_{\ell m_2}a_{\ell m_3}
\left ( \begin{array}{ccc} \ell-1 & \ell+1 & \ell \\ m_1 & m_2 & m_3
\end{array} \right ) \delta_{m_1+m_2+m_3, 0} \nonumber \\ & &
\end{eqnarray}
This procedure can be used to find all the invariants at
any order, within a given multipole ($I^\ell$) and relating
different multipoles ($J^\ell$). These may then be divided by the appropriate
powers of the $C_\ell$ in order to make them dimensionless,
and suitably normalised, as was done with $I^3_\ell$ in \cite{us}.

This method should produce the full set of independent invariant
quantities in a set of $a^\ell_m$. There should be $2\ell -2$
such quantities, for each $\ell$, plus 3 inter-$\ell$ invariants,
for each pair of $\ell$s. The power spectrum measures how much
power there is on a given scale $\ell$. The $I^\ell$ describe
how the power is divided between the various $m$ modes, for a given $\ell$. 
This encodes preferences for shapes within a given multipole. 
The $J^\ell$ measure correlation between the orientations of preferred
shapes in adjacent multipoles.

\section{DMR bispectrum}
\label{chi2}
We now summarise the application of the simplest of these
statistics to DMR.
We will be testing the  inverse noise variance weighted, average maps of 
the 53A, 53B, 90A and 90B {\it COBE}-DMR channels, with monopole
and dipole removed, at resolution 6, in ecliptic 
pixelization. We use the  
extended galactic cut of \cite{banday97}, and 
\cite{benn96} to remove most of the emission from the plane of the Galaxy.
We apply our statistics to the DMR maps before and after correction
for the plausible diffuse foreground emission outside the galactic plane
as described in
\cite{kog96b} and \cite{COBE}. 
To  estimate the $I^3_\ell$s we set
the value of the pixels within the galactic cut to 0 and 
the average temperature {\it of the cut map} to zero. 
We then integrate the
map multiplied with spherical harmonics  to obtain the estimates of the
$a_{\ell m}$s and compute the $I^3_\ell$ from these.
\begin{figure}
\centerline{\psfig{file=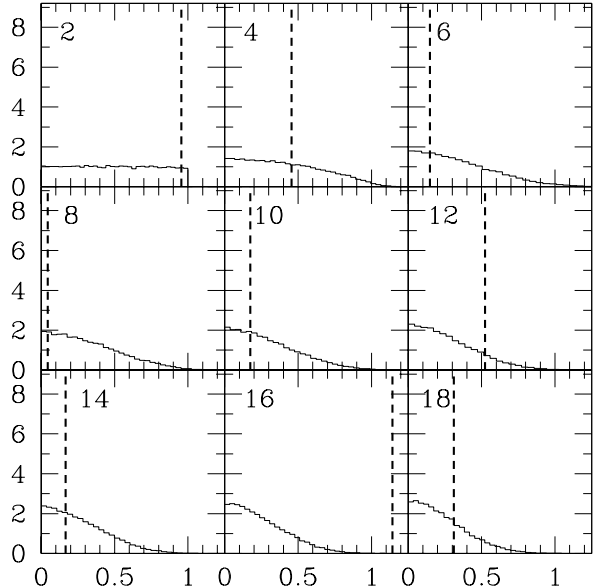,width=8cm}}
\caption{Fig.1 The vertical thick dashed line represents the value 
of the observed
$I^3_\ell$.  The solid line is the probability distribution function
of $I^3_\ell$ for a Gaussian sky with extended galactic cut and
DMR noise.}
\end{figure}

We have used Monte Carlo simulations to find the distribution 
of the estimators $I^3_\ell$ as applied to Gaussian maps subject to
DMR noise and galactic cut (see Fig.1). These distributions
are very non-Gaussian.  In principle this would complete 
the theoretical work required for converting the observed $I^3_\ell$ 
(which we also plot in Fig.1) into a statistical 
statement on Gaussianity, but we proceed further by
defining  a new ``goodness of fit'' statistic as follows.

We constructed a tool similar to the $\chi^2$ (often 
used for comparing predicted and observed $C_\ell$ spectra) 
but adapted to the  non-Gaussian distributions $P(I^3_\ell)$:
\begin{equation}\label{presc}
X^2={1\over N}{\sum_\ell X_\ell^2}=
{1\over N}{\sum_\ell (-2\log P_\ell(I^3_\ell) 
+ \beta_\ell),}
\end{equation}
where the constants $\beta_\ell$ are defined so that for each term
of the sum $\langle X_\ell^2\rangle=1$. The definition reduces
to the usual $X^2$ for Gaussian $P_\ell$. 

This is a suitable definition. As an illustration let 
us first approximate the distributions of the $I_\ell^3$
by $P(I^3_\ell)=2(1-I^3_\ell)$ --- 
a good approximation for $\ell$ around 10.
Then $X^2=-2\log(1-I^3_\ell)$. 
Like the standard $X^2$ one has $0<X^2\ll 1$ for observations close
to the peak of the distribution, here at $I^3_\ell=0$. Indeed
$X^2(0)=0$. However the peak of $P(I^3_\ell)$ is far from its average, 
and so the standard $X^2$ would produce $X^2=0$ at the wrong observation.
For observations far 
from the peak of the distribution (but subject to the constraint
$I^3_\ell\le 1$) $X^2$ goes to infinity. In contrast the standard
$X^2$ would always remain finite.

The proposed $X^2$ therefore does for these non-Gaussian distributions 
what the usual $X^2$ does for normal distributions. 
We build a $X^2$ for the {\it COBE}-DMR data by means of Monte
Carlo simulations. We proceed as follows. First we compute the 
distributions $P(I^3_\ell)$, for $\ell=2,...,18$, 
for a Gaussian process as measured subject to our galactic
cut, and pixel noises. These $P(I^3_\ell)$ were inferred 
from 25000 realizations (see Fig.1). 
From these distributions we then build 
the $X^2$ as defined above, 
taking special care with the numerical
evaluation of the constants $\beta_\ell$. We call 
this function $X^2_{COBE}$.
We then find its distribution $F(X^2_{COBE})$
from 10000 random realizations.  This is very well approximated by 
a $\chi^2$ distribution with 12 degrees of freedom. If all $P(I^3_\ell)$
were as in the analytical fit above, we could conclude that we
successfully measured an effective number of useful invariants
equal to 6. This is less than the number of invariants we actually
measured (10) and this is simply due to 
anisotropic noise and galactic cut. However, had
 we used a standard $\chi^2$ statistics
the effective number of useful invariants would be only 3.

We then compute $X^2_{COBE}$ with the actual observations and find
$X^2_{COBE}=1.81$. One can compute $P(X^2_{COBE}<1.7)= 0.98$.
Hence, it would appear that we can
reject Gaussianity at the $98\%$ confidence level.

\section{Systematics}
\label{systematics}
Given the nature of this result checking for systematics has
been the central aspect of our work. We checked a large number
of effects, related to foreground emissions, pixelization effects,
spurious offsets induced by the galactic cut, the cut itself,
the underlying power spectrum, etc, etc. These checks  are reported
in \cite{us,usnew}. 

It is important to stress that the confidence
level quoted above is a lower limit. It corresponds to the worse
case obtained, within the systematic space surveyed. It assumes
a conspiracy theory with all systematics lined up so as to
take the blame for the observed non-Gaussianity. 
If one does not consider this worst case scenario, but takes other
data sets differently treated for systematics, in most cases 
we obtained a confidence level for rejection in excess of 99.5\%.

\begin{figure}
\centerline{\psfig{file=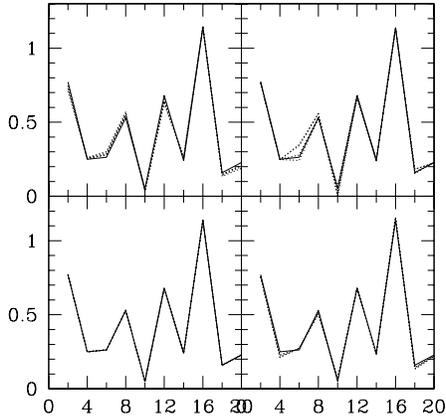,width=6cm}}
\caption{Fig.2 - The effect of adding and subtracting templates of
systematics due the instrument magnetic susceptibility, in directions
$\beta_T$, $\beta_R$, and $\beta_X$, and the effect of potential unknown
effects at spin frequency.}
\label{sys1}
\end{figure}

Checking for systematics is always open ended,
and in these Proceedings we merely present one new 
check we have recently performed. 
It concerns the so-called systematic templates.
The procedure leading from the time series to a CMB map produces,
as well as the map, an estimate of the systematic effects possibly
plaguing the final product. These ``systematic templates'' for DMR
are well documented (\cite{kog96c}), and display strongly
non-Gaussian structures, tracing the DMR scanning patterns.

The systematic templates have negligible power. The worst effect in
the worst channel has a rms of about 6 $\mu K$ at the $95\%$
confidence level. It is unlikely that these effects could corrupt
power spectrum estimates. Nonetheless it is well known that a 
non-Gaussian pattern with negligible power may visually stand out over  
a Gaussian map with much larger power. Similarly it could
happen that these systematics, while irrelevant for 
the purpose of power spectrum estimation, could be responsible
for the observed nonGaussian bispectra, derived from DMR maps.

In order to address this problem, we subjected the systematic
templates to two tests. Firstly we computed the $I^3_\ell$ spectra for 
the templates. The resulting $I^3_\ell$ are well outside the Gaussian 
prediction, but they do not correlate with the DMR observed $I^3_\ell$.
Secondly, we added or subtracted these templates enhanced by a
factor of up to 4 to DMR maps. The effect on the $I^3_\ell$ spectrum
was always found to be negligible. This shows that 
the systematic effects documented in \cite{kog96c} have not only
negligible power, but also negligible effect on the 
higher order statistics which we have studied. 

To be more specific we have applied the above tests to templates for 
systematics in $53A$, $53B$, $90A$, and $90B$, separately. 
This is the sensible thing to do, given that the templates
are highly correlated from pixel to pixel(\cite{kog96c}). 
We have considered the effect of instrument susceptibility to the Earth 
magnetic field; any unknown effects at the spacecraft spin period;
errors in the calibration associated with long-term drifts, and 
calibration errors at the orbit and spin frequency; 
errors due to incorrect removal of the COBE Doppler and  Earth Doppler
signals; errors in correcting for  emissions from the Earth, and
eclipse effects; artifacts due to uncertainty in the correction for
the correlation created by the low-pass filter on the
lock-in amplifiers (LIA) on each radiometer;
errors due to emissions from the moon, and the planets. 
Presumably what is usually mentioned as De-striping goes under 
the removal of artifacts in the calibration. DMR did not seem to have
a serious striping problem, but the problem was addressed none the less.

In Fig.2 we plot the result of adding templates associated with effect
of the magnetic field of the Earth. The instruments' magnetic 
susceptibility, and the emissions from the Earth have 
by far the strongest effects.
The effect on the $I^3_\ell$ spectrum is always negligible, but when 
present occurs at scales around
$\ell=6$.

\section{Cautionary remarks and a digression}
By now two other groups have reported results similar to
ours, albeit making use of different methods \cite{nov,pando}. 
According to skeptics, this may merely reflect a change in the 
psychological prior, triggered by our work. More seriously
one should remember that the work performed by us and by these
groups makes use of the same data set. Therefore this work 
provides an independent confirmation of our analysis of the DMR
maps, but not an independent confirmation of the result itself.
In particular we feel that the issue of systematics, and foreground
contamination, will only 
be  clarified further when an independent all sky data set becomes 
available. 

If indeed our result is due to cosmic emission, then a number
of fascinating theoretical issues are raised (see \cite{nvs}).
 Clearly the minimal inflationary models
cannot be right. On the other hand it is not obvious that 
the main competitor to inflation, topological defects, could 
explain this type of non-Gaussianity. Topological defects 
are non-Gaussian, but in ways which are often more
subtle than commonly thought. Computing with defects is prohibitively
expensive, and  predicting a set of $I^3_\ell$ distributions
in defect models is well beyond current computer technology.

An interesting possibility was recently proposed by Peebles
\cite{peeb}. This is an isocurvature model in which the underlying
fluctuations are not a Gaussian random field, but the square 
of a Gaussian random field. The model is based on 
non minimal inflation, but produces fluctuations radically different
from minimal inflationary fluctuations. The main advantage of this
model over defects 
is that, while not trivial, it is easy enough to compute with it.
In particular it is feasible, and topical, to repeat the exercise
we have performed in \cite{us} using Peebles theory. This model
could well produce
a better fit to the DMR bispectrum than Gaussian theories.

\bsk
\baselineskip = 12pt
{\abstract \ni ACKNOWLEDGEMENTS
We thank Al. Kogut for supplying the systematic maps. JM thanks
the organizers for an excellent meeting, and a great time.
PGF was supported by NSF (USA), JNICT (PORTUGAL) and CNRS (FRANCE),
JM by the Royal Society and TAC, and KMG by Danmarks
Grundforskningsfond (TAC) and partly by NASA-ADP grant.
}

\bsk
\baselineskip = 12pt

% beginning of font "references"

\end{document}